\documentstyle[psfig]{mn}

\input{epsf}

\begin{document}

\title{Determination of the linear mass power spectrum from the mass function of
galaxy clusters}
\author[A. G. S\'anchez, N. D. Padilla \& D. G. Lambas]
{Ariel G. S\'{a}nchez$^1$, Nelson D. Padilla $^{1,2}$ and Diego G. Lambas$^{1,3,4}$ \\
$^1$ Grupo de Investigaciones en Astronom\'{\i}a Te\'{o}rica y Experimental,
 (IATE), Observatorio Astron\'{o}mico C\'{o}rdoba, UNC, Argentina.\\
$^2$ Dept Physics, University of Durham, South Road, Durham, DH1 3LE.\\
$^3$ Consejo Nacional de Investigaciones Cient\'{\i}ficas y Tecnol\'ogicas 
(CONICET), Argentina.\\
$^4$ John Simon Guggenheim Fellow.\\}

\date{Accepted 2002 July 23. Received 2002 July 19; in original form 2002
June 25}
\maketitle

\begin{abstract}
We develop a new method to determine the linear mass power spectrum using
the mass function of galaxy clusters. We obtain the rms mass fluctuation
$\sigma (M)$ using the expression for the mass function in
the Press \& Schechter (1974),
Sheth, Mo \& Tormen (2001) \& Jenkins et al. (2001) formalisms.
We apply different techniques to recover the adimensional power spectrum
$\Delta ^{2}(k)$ from $\sigma (M)$
namely the $k_{eff}$ approximation, the singular value decomposition and the
linear regularization method.
The
application of these techniques to the $\tau $CDM and $\Lambda $CDM GIF
simulations shows a high efficiency in recovering the theoretical
power spectrum
over a wide range of scales. 
We compare our results with those derived from the power spectrum
of the spatial distribution 
of the same sample of clusters in the simulations obtained by application of 
the classical Feldman, Kaiser \& Peacock (1994), FKP, method.
We find that the mass function based method presented here
can provide a very accurate estimate of the linear
power spectrum, particularly for low values of $k$. This estimate is 
comparable, or even better behaved, than the FKP solution. 

The principal advantage of our method 
is that it allows the determination of the linear mass power spectrum
using the joint information of objects of a wide range of masses 
without dealing with specific assumptions on the bias relative to the 
underlying mass distribution.
\end{abstract}


\section{Introduction}

\noindent Cosmological models predict the large scale matter distribution
evolved from primordial fluctuations in the density of the
universe. Unfortunately this distribution is not directly accessible through
observations because we can only observe the distributions of objects like
galaxies or clusters of galaxies that do not, necessarily, trace the mass
distribution in a simple way.

The power spectrum of the galaxy distribution 
has proven to be the most popular statistic used to characterize the
matter distribution in the universe.  This is due to its tight relation
to physical properties of the growth of structures throughout the history
of the Universe.  Also, the many ways and different kind of data that
can be used to measure the power spectrum has
allowed independent results from redshift and angular galaxy
surveys, cluster surveys, and CMB fluctuations studies.
Each of these approaches have different drawbacks, though.  For instance,
the results 
from large redshift surveys, are subject to several problems 
since the results obtained are affected by the
incompleteness of the survey, its geometry, the redshift space distortions
and the bias factor (which can be scale dependent). Moreover the
distribution of galaxies or clusters of galaxies 
is affected by non-linear evolution
effects, which makes it even more difficult to carry out a direct comparison 
between observational
results and theoretical model predictions.

An alternative technique was presented by Gazta\~{n}aga \& Baugh (1998), and
has been applied in different ways by several other authors ( Dodelson \&
Gasta\~{n}aga, 2001, Eisenstein \& Zaldarriaga, 2001 
and references therein). In it the power spectrum of the galaxy distribution
is derived from the angular correlation function inverting Limber's
equation. As this procedure does not involve any Fourier transforms nor the
use of redshift data, the results obtained in this way are not affected by a
convolution with the survey 
window function, nor by redshift space distortions. The
principal uncertainty in this approach comes from the fact that one of 
the ingredients of the technique is the 
redshift distribution of galaxies, which may not be accurately known,
and moreover the
results are still affected by the bias factor and non-linear evolution.

The aim of this work is
to develop a new method for obtaining the power spectrum, free of the
problems of the methods previously described.  In order to do this, 
we use the formalisms of Press \& Schechter (1974,
hereafter PS) and Sheth, Mo \& Tormen (2001, hereafter SMT) 
where halos form in the higher peaks of the evolving primordial 
density fluctuations which are related to the linear power
spectrum.
This indicates that studying the
distribution of halo masses, one can obtain information about the mass 
distribution, and therefore,  
recover the mass power spectrum. 
More precisely, PS's like recipes are sets of equations
that allow the determination of the mass function of dark matter halos from
the mass power spectrum. If the mass function can be determined 
observationally, we can use these equations in the inverse way to derive 
the linear power spectrum.

In this paper we present a new method for determining the linear power spectrum
from the mass function of clusters of galaxies based on the PS, SMT and
Jenkins et al. (2001, hereafter J01) prescriptions for the mass function of
systems expected in hierarchical clustering scenarios. 

The outline of the
paper is as follows, in section \S 2 we briefly review the different
formalisms for obtaining a mass function 
used in this work. In \S 3 we give the general outline of our
method and a simple test in an idealized case. In \S 4 we describe a test of
our method by its application to the GIF simulations ( Kauffman et al.,
1999). In \S 4 we show a comparison with the results obtained using the
standard technique of Feldman, Kaiser \& Peacock (1994, hereafter FKP)  
for measuring the power spectrum, and finally, 
in \S 5 we present a short discussion and the main
conclusions.

\section{The Mass Function}

An important tool with which 
to characterize the density field is the rms density 
fluctuation,
smoothed on some comoving scale $R$.  For the case of a top-hat window,
this scale is 
related to a mass $M=\frac{4\pi }{3%
}\bar{\rho}R^{3}$, where $\bar{\rho}$ is the mean density of the universe ($%
\bar{\rho}=\rho _{c}\Omega $). In turn, this quantity and the power
spectrum $P(k)=\left| \delta _{k}\right| ^{2}$ are related by 
\begin{equation}
\sigma (R)=\int_{-\infty }^{\infty }\Delta ^{2}(k)W^{2}(kR)\ {\rm d}\ln k
\label{sigmaydelta}
\end{equation}
where $\Delta ^{2}(k)=\frac{k^{3}}{2\pi ^{2}}P(k)$ is the dimensionless power
spectrum and $W(kR)$ is the Fourier transform of the top-hat window function 
\[
W(kR)=\frac{3}{kR}j_{1}(kR)=3\frac{(\sin (kR)-kR\cos (kR))}{(kR)^{3}}, 
\]
where $j_{1}(x)$ is the spherical Bessel function of order 1.

\begin{figure}
{\epsfxsize=8.truecm \epsfysize=8.truecm 
\epsfbox[20 150 580 710]{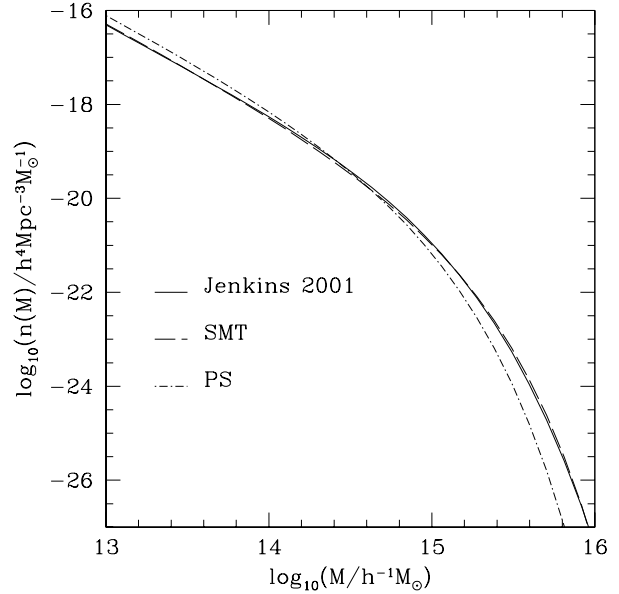}}
\caption{
The mass functions predicted
by different formalisms for a $\protect\tau $CDM model. The curve
obtained from the SMT formalism gives a much better description of the
results from numerical simulations (shown by 
the J01 general fit) than the PS recipe.
}
\label{fig1}
\end{figure}

PS developed a simple formalism to obtain an analytical expression for the
differential mass function of dark matter halos based on the spherical
collapse model. In this model, the abundance of haloes with masses
$M \rightarrow M+{\rm d}M$ is given by: 
\begin{eqnarray}
n(M){\rm d}M &=&\left( \frac{2}{\pi }\right) ^{\frac{1}{2}}\frac
{\bar{\rho}}{M^{2}}%
\frac{\delta _{c}}{\sigma (M)}\left| \frac{d\ln (\sigma (M))}{d\ln (M)}%
\right|  \label{ps} \\
&&\times \exp \left[ -\frac{\delta _{c}^{2}}{2\sigma (M)^{2}}\right] {\rm d}M, 
\nonumber
\end{eqnarray}
where $\delta _{c}$ is the density threshold corresponding to collapse
according to the spherical collapse model. For an $\Omega =1$ universe $%
\delta _{c}=1.686$ and its value has little dependence on cosmology (Eke et
al., 1996). This function is plotted in Figure 1 for a $\tau $CDM model.

The vague analytical justification in the derivation of expression (\ref{ps}%
) has lead to different criticisms of the PS formalism ( e.g. Peacock \&
Heavens, 1989), but a later and more formal derivation due to Bond et al. (
1991) based on a particular type of filtering, and 
the set excursion theory, yields the same expression for the
mass function according to the spherical collapse model, placing the PS
formalism on a more solid theoretical basis. In addition to the
original tests presented in PS, 
equation (\ref{ps}) has been checked against results
from a variety of numerical simulations (Efstathiou et al., 1988, Colberg \&
Couchman, 1989) showing a good general agreement.

Nevertheless, recent works (Tormen, 1998; Sheth \& Tormen, 1999, Governato
et al. 1999, J01)
detected discrepancies between the PS predictions and the halo abundances
obtained in much better numerical simulations. The PS formula underestimates
 the abundances of massive halos and overestimates the low mass end of the
 mass function (already seen in the earlier numerical simulations results). 
A better description of the numerical results is given by a
correction to the PS mass function proposed originally by Sheth \& Tormen
(1999)

\begin{eqnarray}
n(M){\rm d}M &=&\sqrt{\frac{2aA^{2}}{\pi }}\frac{\bar{\rho}}{M^{2}}
\frac{\delta_{c}}{\sigma (M)}\left[ 1+\left( \frac{\sigma (M)}
{\sqrt{a}\delta _{c}}%
\right) ^{2p}\right]  \label{smt} \\
&&\times \left| \frac{d\ln \left( \sigma \right) }{d\ln (M)}\right|
e^{-\left( \frac{a\delta _{c}^{2}}{2\sigma ^{2}(M)}\right)}{\rm d}M
\end{eqnarray}
where $A=0.322$, $a=0.707$ and $p=0.3$. This function is also plotted in
Figure 1. SMT showed that this correction can be understood as a result of
the incorporation of the ellipsoidal rather than spherical collapse model
dynamics in a formalism similar to the one used by Bond et al. (1991).

We can define a function
\begin{equation}
f(\sigma ,z)=\frac{M}{\bar{\rho}}\frac{dn_{cum}(M,z)}{d\ln \sigma ^{-1}},
\label{fdes}
\end{equation}
where $n_{cum}(M,z)$ is the cumulative mass function. From (\ref{ps}) and (%
\ref{smt}) we can calculate $f(\sigma ,z)$ for PS and SMT formalisms where
\begin{eqnarray}
f(\sigma ;PS) &=&\sqrt{\frac{2}{\pi }}\frac{\delta _{c}}{\sigma }\exp \left(
-\frac{\delta _{c}^{2}}{2\sigma ^{2}}\right)  \label{fdesigmaPS} \\
f(\sigma ;SMT) &=&\sqrt{\frac{2a}{\pi }}\left[ 1+\left( \frac{\sigma ^{2}}{%
a\delta _{c}^{2}}\right) ^{p}\right] \frac{\delta _{c}}{\sigma }\exp \left( -%
\frac{a\delta _{c}^{2}}{2\sigma ^{2}}\right)  \label{fdesigmaSMT}
\end{eqnarray}
That is, the PS and SMT formalisms predict that the function $f(\sigma ,z)$ is
independent of redshift and the cosmological model when it is expressed in
terms of $\sigma$.

From the analysis of a set of simulations carried out by the Virgo
Consortium, J01 found 
that the function $f(\sigma )$ can be described with an accuracy
better than 20\% for different redshifts and initial conditions
by the general fit 
\begin{equation}
f(\sigma )=A\exp \left( -\left| \ln \sigma ^{-1}+B\right| ^{\varepsilon
}\right),  \label{jenkinsf}
\end{equation}
where $A=0.315$, $B=0.61$ and $\varepsilon =3.8$, for halos
identified with the FOF algorithm using the same linking
length $b=0.2$ for every value of $\Omega$.

The expression for $n(M)$ inferred from (\ref{fdes}) and (\ref{jenkinsf}) 
\begin{equation}
n(M,z){\rm d}M=\frac{A\bar{\rho}(z)}{M^{2}}\frac{d\ln (\sigma ^{-1})}
{d\ln (M)}%
e^{-\left( \left| \ln (\sigma ^{-1}+B\right| ^{\varepsilon }\right) }
{\rm d}M
\label{j01}
\end{equation}
This differential mass function is also shown in Figure 1, where it can be
seen that SMT formalism is in a much better agreement with the general fit
of J01 (that is, the results of numerical simulations), than PS formalism.

\section{The Method}

In this section we analyse the problem of recovering the power
spectrum $\Delta ^{2}(k)$ using equation (\ref{sigmaydelta}) and the
expressions for the mass function in the formalisms just described.
The steps to be followed in obtaining the linear mass power spectrum are:
\begin{enumerate}
\item Obtain the variance $\sigma(M)$ from the mass function.  In order
to do this we need to rewrite the expresions relating these statistics.
\item Invert the power spectrum from the variance.  This involves the
application and test of different inversion algorithms.
\end{enumerate}

\subsection{Obtaining $\protect\sigma (M)$ from $n(M)$}

Assuming the mass function of dark matter haloes obeys
one of the three formalisms described in the last section, 
we can re-interpret equations (\ref{ps}), (\ref{smt}) and (\ref{j01}) as
differential equations and use them in 
the determination of $\sigma (M)$ provided
$n(M)$ is known. In the Appendix we describe the solution of the equations
obtained for each formalism.

In the case of PS formalism the final solution is

\begin{equation}
\sigma (M)=\frac{\delta _{c}}{\sqrt{2}{\rm erf}^{-1}\left[ {\rm erf}(\frac{%
\delta _{c}}{\sqrt{2}\sigma _{8}})-\frac{G(M)}{\bar{\rho}}\right] },
\label{sigmadeps}
\end{equation}
where ${\rm erf}(x)$ is the error function and $G(M)$ is given by

\[
G(M)\equiv \int_{M}^{M_{8}}\grave{M}n(\grave{M}){\rm d}\grave{M}. 
\]

For the SMT formalism we get 
\begin{equation}
\sigma (M)=\frac{\delta _{c}}{\sqrt{2}\Phi ^{-1}\left[ \Phi (\frac{\delta
_{c}}{\sqrt{2}\sigma _{8}})-\frac{G(M)}{\bar{\rho}A}\right] },
\label{sigmadesmt}
\end{equation}
where we keep the definition of $G(M)$ and 
\begin{equation}
\Phi (x)={\rm erf}(x)+\frac{2^{-p}}{\sqrt{\pi }}\Gamma \left( \frac{1}{2}%
-p\right) P\left( \frac{1}{2}-p,x^{2}\right),  \label{fismt}
\end{equation}
where $p$ is a parameter in the SMT formalism,
and $\Gamma \left( x\right) $ and $P\left( a,x\right) $ are the gamma %
and incomplete gamma functions.

Finally, for the J01 prescription,
\begin{equation}
\sigma (M)=\exp \left\{ -\Psi ^{-1}\left[ \Psi (\ln \sigma _{8}^{-1})-\frac{%
G(M)}{A\bar{\rho}}\right] \right\},  \label{sigmadej01}
\end{equation}
where the function $\Psi (S)$ is given by

\[
\Psi (S)=\int_{0}^{S}e^{-\left| x+B\right| ^{\varepsilon }}{\rm d}x. 
\]
In every case the values of $\Omega $ and the amplitude of 
density fluctuations measured in a sphere of $8 h^{-1}$ Mpc,
$\sigma_{8}$, are assumed to be known.

If we associate clusters of galaxies with dark matter haloes we can obtain $%
\sigma (M)$ from an observational estimation of the differential mass
function of galaxy clusters, using equations (\ref{sigmadeps}), (\ref
{sigmadesmt}) or (\ref{sigmadej01}), assuming it is well described
by the PS, SMT or J01 formalisms. 
In order to obtain a realistic $\sigma (M)$ we must ensure
that the assumed formalism provides an accurate description
of the underlying mass function. 
The most reliable results will be obtained for the
formalism that best describes the mass function, a fact that can be
tested using numerical simulations. 

\subsection{Inverting $P(k)$ from $\sigma(M)$}

\subsubsection{A First try}

As a result of the procedures described in the last section we obtain a set
of values $\sigma (R_{i})$ for $i=1,...,N_{R}$. For each one of them we have
an equation that relates this variance to the power spectrum $\Delta ^{2}(k)$%
\begin{equation}
\sigma ^{2}(R_{i})=\int_{-\infty }^{\infty }\Delta ^{2}(k)W^{2}
(kR_{i}){\rm d}\ln(k)  
\label{sigmacondelta}
\end{equation}

This is an integral Fredholm equation, with a kernel function given by the
Fourier transform of the window function. With this equation, we are
able to recover an
estimate of $\Delta ^{2}(k)$ in a set of values $k_{\alpha }$ $\alpha
=1,...,N_{k}$. Following Dodelson \& Gazta\~{n}aga (1999) we will use Greek
index for quantities in $k$-space and Latin for those in real-space.

A first and rather simple procedure to achieve this is to use the fact that
equation (\ref{sigmacondelta}) is sharply peaked around a characteristic
value of $k$ and only a relatively small interval of $k$-values 
contributes to this integral. Then, $\sigma^2(R)$ can then be understood as
 an average value of the power spectrum in that interval that satisfies 
the relation ( Peacock, 1991; Padilla \& Baugh, 2001):

\begin{equation}
\sigma ^{2}(R)=\Delta ^{2}(k_{eff})  \label{kefectivo}
\end{equation}
for some suitable defined value of $k_{eff}$.

In fact, if we assume that the power spectrum is a power law $\Delta
^{2}(k)=Ak^{n}$, this relation is exact and $k_{eff}$ is given
by

\begin{equation}
k_{eff}=\frac{1}{R}\left[ 9I(n)\right] ^{\frac{1}{n}},  \label{keff}
\end{equation}
where

\[
I(n)=\int_{0}^{\infty }y^{n-7}\left[ \sin (y)-y\cos (y)\right] ^{2}{\rm d}y. 
\]

It is not necessary to assume that the power spectrum behaves like a power
law on every scale, as it was mentioned earlier. 
This can be seen by considering that for a given value of $R$
only a small range in $k$ is important, and locally we can approximate the
power spectrum by a power law $\Delta ^{2}(k)\propto k^{n}$ and
so equation (\ref{keff}) is expected to hold at least
approximately for this value of $n$.   This is specially true in the case
of CDM class power spectra, which benefit from having slowly varying $n(k)$.

\begin{figure}
{\epsfxsize=8.truecm \epsfysize=8.truecm 
\epsfbox[20 150 580 710]{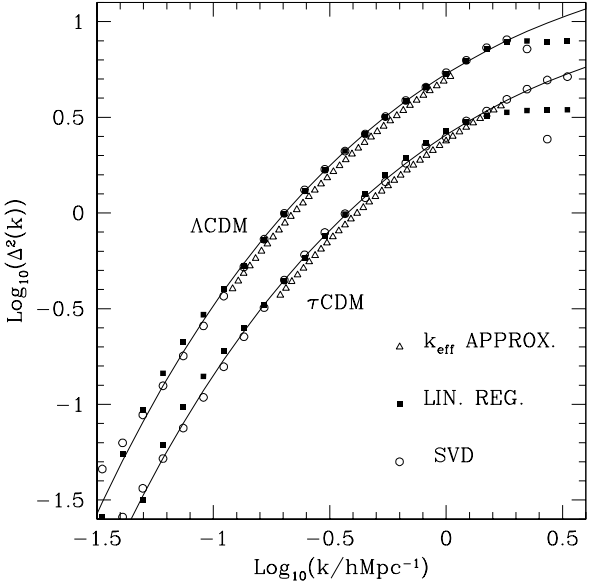}}
\caption{
Results of the
implementation of the $\Delta ^{2}(k_{eff})$ approximation, SVD and linear
regularization to the functions $\protect\sigma (M)$ obtained from the power
spectrum of the $\protect\tau $CDM and  $\protect\Lambda $CDM models
}
\label{fig2}
\end{figure}

We can't use this procedure directly to obtain $\Delta ^{2}(k)$ because
equation (\ref{keff}) depends on the index $n$ of the unknown power
spectrum. We can avoid this difficulty by either assuming a fixed value for $%
n $ on every scale ( e.g. $n=-2$ for scales corresponding to clusters of
galaxies) or by obtaining 
it from $\sigma (M)$. If $\sigma ^{2}(M)$ behaves like a
power law $\sigma ^{2}(R)\propto R^{\tilde{n}}$ and

\[
n=-\tilde{n} 
\]

Figure 2 shows the power spectrum recovered from the function $\sigma (M)$
calculated for a $\tau $CDM and a $\Lambda $CDM power spectrum using this
method. The range of masses used in this calculation is 
$10^{12}h^{-1}M_{\odot }<M< 10^{15}h^{-1}M_{\odot }$. 
It is clear from this figure that
this procedure gives good results in spite of its simplicity. 
The solutions obtained for $\Delta ^{2}(k)$ have the correct
shape but its amplitude is somewhat underestimated.

\subsubsection{Singular Value Decomposition}

If we denote by $D_{\alpha }$ and $S_{i}$ the values of $\Delta ^{2}(k_{\alpha
})$ and $\sigma ^{2}(R_{i})$, 
the integral equation (\ref{sigmacondelta}) can be cast as
a matrix, yielding
\begin{equation}
\mathbf{S=}K\mathbf{D}  \label{matrix}
\end{equation}
where $K$ is the $N_{\sigma }\times N_{k}$ kernel matrix
and the vectors $\mathbf{D}$
and $\mathbf{S}$ are formed by the values $D_{\alpha}$ and $S_i$ respectively. 
Expressed in this
way, the first thought about how to obtain a solution is by direct inversion
of the kernel matrix, that is
\begin{equation}
\mathbf{D}=K^{-1}\mathbf{S}.  \label{inversiondirecta}
\end{equation}
Unfortunately this simple approach does not work. The solutions obtained 
inverting $K$ are unstable and often do not have a true physical meaning. 
This due to the fact that we try to obtain more information about the 
power spectrum than is contained in $\sigma (M)$. 
This situation causes the matrix $K$ to
be numerically singular ( although not \textit{strictly} singular).

\begin{figure}
{\epsfxsize=8.truecm \epsfysize=8.truecm 
\epsfbox[20 150 580 710]{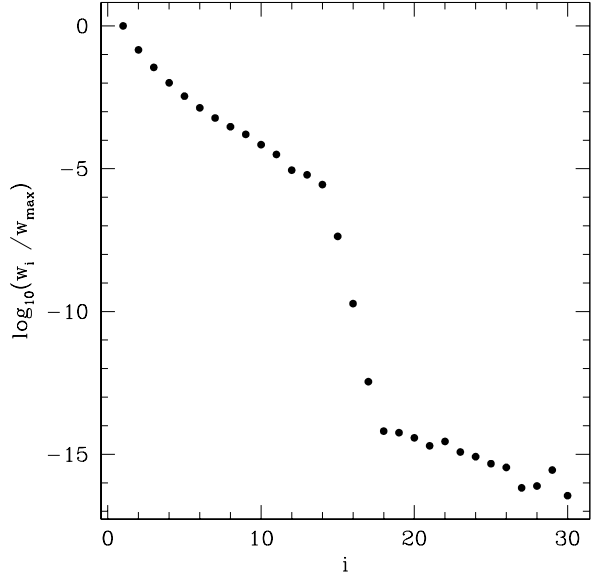}}
\caption{
Value of $\log (\frac{w_{i}%
}{w_{\max }})$ for the different singular values of the matrix $\mathbf{W}$.
A great quantity of the $w_{i}$ are very small and are susceptible to
roundoff errors.
}
\label{fig3}
\end{figure}
In order to obtain 
the best solution $\mathbf{D}$ it is necessary to assume that the errors in
$S_{i}$ obtained from the mass function have a gaussian distribution
around their true values $\bar{S}_{i}$, then the probability to obtain a
vector $\mathbf{S}$ is proportional to $\exp (-\frac{\chi ^{2}}{2})$ where 
\begin{equation}
\chi ^{2}=\left( \mathbf{S-\bar{S}}\right) ^{T}C_{S}^{-1}\left( \mathbf{S-%
\bar{S}}\right)  \label{chicuadrado},
\end{equation}
and $C_{S}$ is the covariance matrix for $\mathbf{S}$.

We can relate $\mathbf{\bar{S}}$ to the true value of the power spectrum by $%
\mathbf{\bar{S}=}K\mathbf{D}$. It is easy to see that the solution $\mathbf{D}$
that minimizes (\ref{chicuadrado}) is simply $\mathbf{D}=K^{-1}\mathbf{S}$,
and that its covariance matrix will be given by $C_{D}=\left(
K^{T}C_{S}^{-1}K\right) ^{-1}$, but, as we mentioned before, this solution
does not work. To solve this problem we use the Singular Value
Decomposition technique (SVD, see Press et al. 1992). Following Eisenstein
\& Zaldarriaga (2001), we define $\tilde{K}=C_{S}^{-\frac{1}{2}}K$ and $%
\mathbf{\tilde{S}=}C_{S}^{-\frac{1}{2}}\mathbf{S}$, where $C_{S}^{-\frac{1}{2%
}}$ is the inverse matrix of $C_{S}^{\frac{1}{2}}$ constructed by taking the
square root of the eigenvalues of $C_{S}$. We then have 
\[
\chi ^{2}=\left| \tilde{K}\mathbf{D}-\mathbf{\tilde{S}}\right| ^{2}. 
\]

SVD can be used to find the solution $\mathbf{D}$ that minimizes
this quantity. The singular value decomposition of the $\tilde{K}$ 
matrix is given by $\tilde{K}=UWV^{T}$, where $U$ is a 
$N_{R}\times N_{k}$ column orthogonal matrix, $W$ is an 
$N_{k}\times N_{k}$ diagonal matrix whose entrances $w_{i}$ are 
called singular values, and $V$ is an $N_{k}\times
N_{k}$ orthogonal matrix. With this notation we can express the solution (%
\ref{inversiondirecta}) as $\mathbf{D}=VW^{-1}U^{T}\mathbf{S}$, and the
covariance matrix $C_{p}=VW^{-2}V^{T}$. If the kernel matrix is singular,
then some of the elements $w_{i}$ of the matrix W 
are zero and the inverse matrix $W^{-1}$ is obviously not defined. But, even
when none of the singular values are exactly zero, their values can be so
small as to be dominated by the roundoff errors and their inverses $%
w_{i}^{-1}$ will have very large values causing the problem to be
numerically intractable. Figure 3 shows the singular values $w_{i}$ obtained
by performing the SVD of the matrix $K.$ It can be seen that there are a large
number of singular values that are very small causing the problem to be
singular.

The solution obtained replacing
by zero the quotients $w_{i}^{-1}$ corresponding to the smallest singular
values minimizes equation (\ref{chicuadrado}) and is therefore the solution $%
\mathbf{D}$ that we are looking for
( Press et al., 1992).

In Figure 2 is also shown 
the power spectrum recovered from $\sigma (M)$ 
by application of the SVD method 
for the same range of masses than in the 
$\sigma ^{2}(R)=\Delta ^{2}(k_{eff})$ approximation. It is clear from this
figure that SVD gives very good results in a wide range of scales,
correctly reproducing the shape and amplitude of the theoretical
power spectrum.

\subsubsection{Linear Regularization}

Another way to obtain a solution for (\ref{matrix}) is using a Bayesian
approach. As we mentioned before the problem we are trying to solve is
singular since we wish to obtain more information about $\Delta ^{2}(k)$
from $\sigma (M)$ than that actually available. To avoid this lack
of information we can supply a prior such as the hypothesis of a
``smooth''  power spectrum. In order to do this we put a
second exponential in the probability distribution for $\mathbf{S,}$ which
will be proportional to $\exp \left[ -\frac{1}{2}\left( \chi ^{2}+\lambda
\beta \right) \right] $, where 
\begin{equation}
\beta =\mathbf{D}^{T}H\mathbf{D}.  \label{beta}
\end{equation}

Now we have to find the solution $\mathbf{D}$ that minimizes the weighted
sum $\chi ^{2}+\lambda \beta $. If the minimization of the functional $\beta 
$ alone is non-singular, the general minimization will be so too (Press et
al., 1992). The first functional ( $\chi ^{2}$) is a measure of the
concordance of the solution $\mathbf{D}$ with the data $\mathbf{S}$, the
second term ( $\beta $) can be viewed as a measure of its agreement with a 
\textit{prior,} implemented by the matrix $H$,
that makes $\mathbf{D}$ not to vary too much. The $\lambda $
factor in the second term in eq. (\ref{beta}) sets the relative weight of the
two functionals. The limit $\lambda \rightarrow 0$ correspond to the
minimization of $\chi ^{2}$. In this work we have chosen the matrix $H$
corresponding to a constant $\mathbf{D}$ (equation 18.5.3 in Press et al.,
1992).

With these definitions the solution for $\mathbf{D}$ that minimizes $\chi
^{2}+\lambda \beta $ can be found by solving the \textit{normal equations} 
\[
\left( \tilde{K}^{T}\tilde{K}+\lambda H\right) \mathbf{D}=\tilde{K}^{T}%
\mathbf{\tilde{S}} 
\]
and the covariance matrix will be given by 
\[
C_{D}=\left( \tilde{K}^{T}\tilde{K}+\lambda H\right) ^{-1}. 
\]

The solutions obtained in this way depend on the value of $\lambda $.
This must be determined according to our confidence in observational data. A
value that gives equal weight to the minimization of the functionals $\chi
^{2}$ and $\beta $ is given by 
\begin{equation}
\lambda ={\rm Tr}(\tilde{K}^{T}\tilde{K})/{\rm Tr}(H).
\label{lambdaigualdad}
\end{equation}

The results of the application of this method can be appreciated in
figure 2 where we
show the power spectrum recovered from the same $\sigma (M)$
used in $K_{eff}$ and SVD methods.
The value of $\lambda $ adopted is that of (\ref{lambdaigualdad}). 
It is clear from this figure that Linear Regularization (LR) also 
provide suitable results for $\Delta ^{2}(k)$ in a wide range of scales.

\section{Testing the Method}

\subsection{The GIF simulations}

\begin{figure}
{\epsfxsize=8.truecm \epsfysize=8.truecm 
\epsfbox[20 150 580 710]{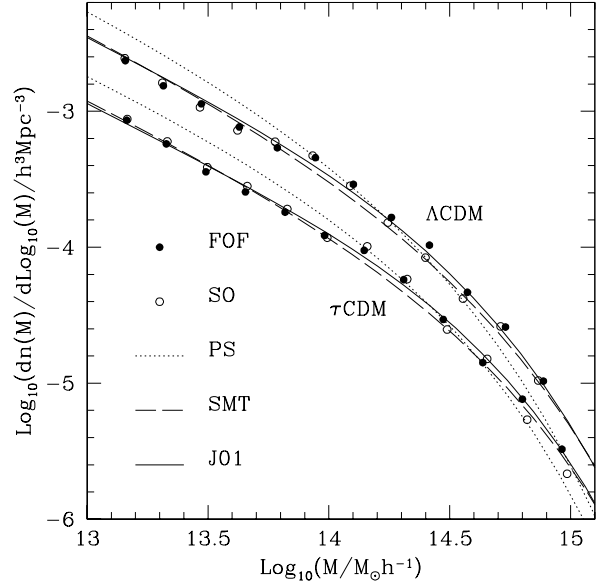}}
\caption{
Mass function from the GIF $%
\protect\tau $CDM and $\Lambda $CDM simulations, determined using FOF halo
finder compared with the predictions of the PS, SMT and J01 formalisms.
}
\label{fig4}
\end{figure}

To test the capability of the method developed in the last section to
recover the true power spectrum from the differential mass function, we
applied it to the GIF $\tau $CDM and $\Lambda $CDM simulations (Kauffman et
al., 1999). The values of the cosmological parameters for each simulation
are listed in Table 1.

In order to apply our method we first determine the mass
function of clusters identified in 
the numerical simulations. We used a Friends Of Friends
(FOF) and a Spherical Overdensity (SO) halo finders. The mass function 
determined in this way is sensitive to the choice of the linking length
parameter $b$ and density contrast $\kappa $ in the FOF and SO 
algorithm respectively. 
Both parameters were set in accordance with
the spherical 
collapse model as described in Eke et al. (1996): $b=0.2$ and 
$\kappa =180$ for $\Omega =1,$ and $b=0.164$ and $\kappa =324$ for 
$\Omega =0.3$.

{\small Table 1: values of the cosmological parameters in the GIF
simulations (Kauffman et al., 1999)}

\begin{tabular}{|l|l|l|l|l|l|c|}
\hline
Model & $\Omega _{o}$ & $\Omega _{\Lambda }$ & $h$ & $\sigma _{8}$ & 
$\Gamma $ & Box \\ \hline
$\tau $CDM & 1.0 & 0.0 & 0.5 & 0.6 & 0.21 & \multicolumn{1}{|l|}{85$h^{-1}%
{\rm Mpc}$} \\ \hline
$\Lambda $CDM & 0.3 & 0.7 & 0.7 & 0.9 & 0.19 & \multicolumn{1}{|l|}{140$%
h^{-1}{\rm Mpc}$} \\ \hline
\end{tabular}

The differential mass functions obtained are plotted in Figure 4,
where we also plot the predictions from the formalisms of PS, SMT and J01
for comparison. As it can be seen, 
the results obtained with the two halo finders are in 
excellent agreement.  This allows us to concentrate on
only one halo finding algorithm which we choose to be the FOF algorithm.
The later choice is based in the fact that
this is the algorithm used by J01 in the determination of the values 
given in \S2 for the coefficients in the mass function of equation (\ref{j01}). 
Another point in support of our choice comes from the argument
presented by J01 which states that the universality of the mass function 
is more evident using FOF than SO.
Note that the prediction of the PS formalism fails to reproduce the mass 
functions obtained from the simulations whereas the SMT and J01 
prescriptions give a more accurate result. This is no surprise since the 
GIF simulations where used to obtain the parameters in the expressions 
for the mass function in these formalisms.

\subsection{The determination of $\protect\sigma (R)$:}

\begin{figure}
{\epsfxsize=8.truecm \epsfysize=8.truecm 
\epsfbox[20 150 580 710]{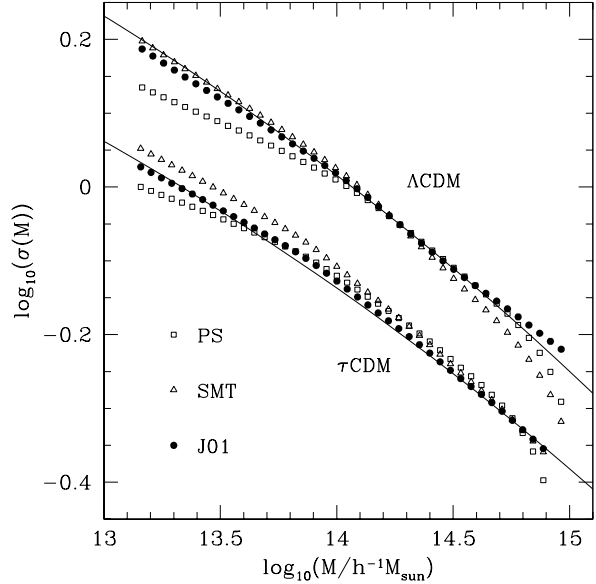}}
\caption{
Solutions for $\protect\sigma %
(M)$ obtained by the application of the different formalisms to the mass
functions of the GIF simulations. For clarity the solutions obtained for 
the $\tau $CDM simulations are divided by a factor 1.25. 
}
\label{fig5}
\end{figure}

The next step in the method described in \S 3 is the determination of $%
\sigma (R)$ from $n(M)$ using equations (\ref{sigmadeps}) (\ref{sigmadesmt})
and (\ref{sigmadej01}). In doing so, we have only used the mass functions
obtained from the simulations for $M>10^{13}$h$^{-1}$M$_{\odot }$, which is
approximately the range that corresponds to clusters of galaxies. Figure 5 shows
the results for $\sigma (M)$ obtained in this way for the $\tau $CDM and $%
\Lambda $CDM simulations, together with the $\sigma (M)$ function
calculated using (\ref{sigmacondelta}) and the power spectrum present in
each model.

The best solutions for $\sigma (M)$ where obtained by using the J01
formalism, indicating that it gives the best description of the mass
function of dark matter halos when these are identified using a FOF\
algorithm.  

The results obtained with the SMT 
formalism are in good agreement
with those obtained using the J01 recipe for the 
$\Lambda $CDM simulation, which
is not the case with the $\tau $CDM simulation, showing that this formalism
does not describe the $\tau$CDM mass function at the level of accuracy
required in our method. 
The results obtained with the PS formalism differ notably
from those obtained from the theoretical model present in the simulations, 
showing
that this represents a poor description of the mass function, a recurrent 
result in the literature on the subject.

We attempted to determine the covariance matrix $C_{S}$ for $\sigma ^{2}(M)$
using a bootstrap resampling technique. We determined $n(M)$ for a large 
number of samples of groups randomly selected from the original list from 
the simulations. For each of them we calculate $\sigma ^{2}(M)$ according 
to the PS, SMT and J01 recipes. Following the notation used in \S 3.2 we 
estimate the covariance matrix by  
\[
(C_{S})_{ij}=\left\langle \left( S_{i}-\bar{S}_{i}\right) (S_{j}-\bar{S}%
_{j})\right\rangle 
\]
where $\bar{S}$ represents the theoretical value of $\sigma ^{2}$ for the
model considered.

\begin{figure}
{\epsfxsize=8.truecm \epsfysize=8.truecm 
\epsfbox[20 150 580 710]{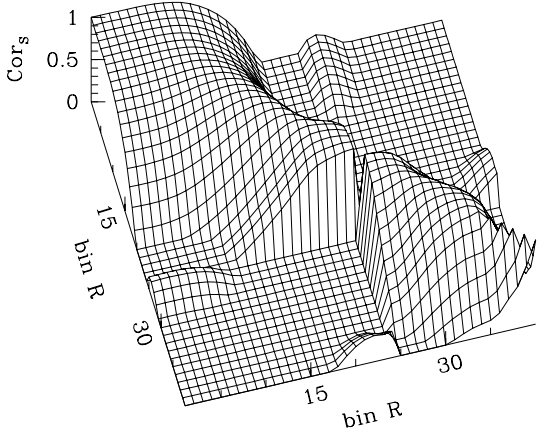}}
\caption{
Correlation matrix of the $%
\protect\sigma ^{2}(M)$ function obtained from the $\Lambda $CDM mass
function using the J01 formalism. This matrix was calculated using a
bootstrap resampling method. 
}
\label{fig6}
\end{figure}

Figure $6$ shows the correlation matrix $Cor_{ij}=\left( C_{S}\right) _{ij}/%
\sqrt{\left( C_{S}\right) _{ii}\left( C_{S}\right) _{jj}}$ obtained
with this procedure
for the J01 solutions determined from the $\Lambda $CDM simulation, the
results from using the other formalisms are similar. 
Note that all masses smaller
(bigger) than $M_{8}$ are correlated. As the mass $M$ only appears in the
expressions for $\sigma (M)$ through the function $G(M)=%
\int_{M}^{M_{8}}Mn(M){\rm d}M$, the value of $\sigma (M)$ for masses lesser
(bigger) than $M_{8}$, have no influence on the value of $\sigma (M)$ for
masses bigger (lesser) than this limit, so the correlations between these two
regions should vanish. This characteristic is not present in the correlation
matrix determined by the bootstrap resampling technique, which shows spurious
correlations that prove the inadequacy of this technique for the
determination of the covariance matrix. We have not succeded in 
determining $C_{S}$ so we assume here a diagonal form. As pointed out
by Eisenstein \& Zaldarriaga (2001) this simplification may cause an
underestimation of the errors in the final power spectrum. 
For the purpose of this work this is a reasonable approximation,
but it has to be properly taken into account if it is aimed 
to obtain precise constraints upon cosmological models using observational 
data.

\subsection{The determination of $\Delta ^{2}(k)$}

Given that in both cases ( $\tau $CDM and 
$\Lambda $CDM) the solution $\sigma (M)$ 
determined using equation (\ref{sigmadej01}) with the general fit from
J01, is the one that shows a better agreement with that 
of the corresponding cosmological model, we use this solution to 
determine the power spectrum assuming that $C_{S}$ is diagonal. Figure 7 
shows the results obtained in this way for the $\tau $CDM and 
$\Lambda $CDM simulations using the different techniques described in 
\S 3.2. For comparison we also plot the theoretical power spectrum 
$\Delta ^{2}(k)$ for each model.

\begin{figure}
{\epsfxsize=8.truecm \epsfysize=8.truecm 
\epsfbox[20 150 580 710]{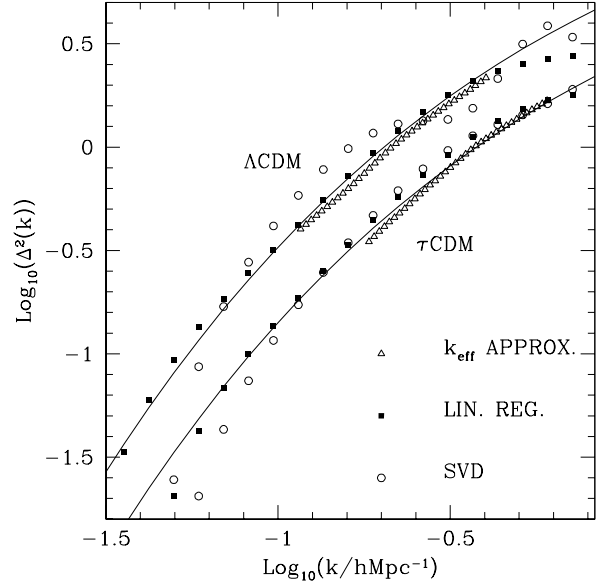}}
\caption{
Solutions for $\Delta ^{2}$%
(k) obtained with the different techniques of \S 3.2, using the solution for 
$\protect\sigma (M)$ determined by de J01 general fit compared with the
correct power spectrum of the $\protect\tau $CDM and $\protect\Lambda $CDM
 models.
}
\label{fig7}
\end{figure}

In the implementation of the approximation $\sigma ^{2}(R)=\Delta
^{2}(k_{eff})$ the value of the index $n$ was not determined from $\sigma
(R) $ as described in \S 3.3.1.  The solution obtained applying this
procedure showed an oscillatory behaviour. Instead, we fixed
$n=-2$, and on those scales where this approximation is valid,
the obtained power spectrum shows a good agreement with that
corresponding to the different models.  The solution starts to fail where 
this approximation ceases to be valid. The solutions recovered have 
approximately the correct shape, but suffer from the same problem 
mentioned in \S 3.2.1, underestimating the amplitude of the power 
spectrum. This analysis shows that this technique 
could represent a powerful tool to determine the power spectrum if 
the index $n$ is determined in a more efficient way.

The solution obtained applying SVD as described in \S 3.3.2 shows a very
good agreement with the correct power spectrum over a wider range of scales
in the $\tau $CDM simulation but fails to reproduce the correct shape for
the $\Lambda $CDM case. This behavior was not present in the solution
plotted in figure 2, where the best results were found using this technique. 
This shows that the solutions obtained using SVD are very sensitive to the
differences between the used and exact $\sigma (M)$, and therefore,
to the deviations of $n(M)$ with respect to the predictions of the J01
general fit.  This indicates that
the implementation of the SVD method 
to real data is not very efficient, a situation that may 
improve with the use of the full covariance matrix, 
instead of a diagonal $C_{S}$.

The best solutions where obtained using the LR
technique, which reproduces the correct shape and amplitude of the power
spectrum over a wide range of scales, showing the same performance that in \S
3.2.3.  This proves that, at least within the hypothesis of a diagonal
covariance matrix, this is the best technique to recover the power spectrum
from the mass function of galaxy clusters.

It should be stressed the fact that the simulations used in this 
test are small in volume.  Therefore we have not been able to analyze 
the capacity of our method to deal with the presence of rare massive 
objects. The formalisms of PS, SMT and J01 have been extensively 
tested against the results of numerical simulations using volumes much 
larger than the ones used in this work. These analyses have 
shown that the SMT and J01 mass functions account properly for the
most massive objects. This indicates that
the effect of the presence of massive clusters
in the mass function will be in perfect agreement with the 
theoretical formalisms. Further more, the inclusion of these objects in 
our analysis will improve our results and increase the 
range of scales in which $\Delta ^{2}(k)$ is recovered, since the numerical 
problem becomes more easily tractable when a wider range of masses
is taken into account.        
                              
To test our method in a more realistic situation,
we also analyzed the effect of introducing errors 
in the determinations of individual cluster masses. 
We constructed $5$ samples of groups from the original one 
identified in the $\Lambda$CDM GIF simulation,
by adding gaussian errors to the individual masses of amplitudes 
10\%, 20\%, 30\%, 40\% and 50\%.
For each of these new samples we determined the 
mass function. The results obtained show that even an error
as large as 50\% in the individual mass 
determinations produces a small overall effect 
in the low mass end, causing deviations of less 
than 10\% in all cases. In the high mass end ( log$_{10}(M/h^{-1}M_{\odot})
 > 14.5$) these errors can produce deviations of even 30\% due to the 
small number of objects. Nevertheless, their presence does 
not produce a considerable effect in the final power spectrum, 
indicating that our method is robust even when dealing with large errors 
in the mass estimates. 

\subsection{Dependence on $\Omega $ and $\protect\sigma _{8}$}

\begin{figure}
{\epsfxsize=8.truecm \epsfysize=8.truecm 
\epsfbox[20 150 580 710]{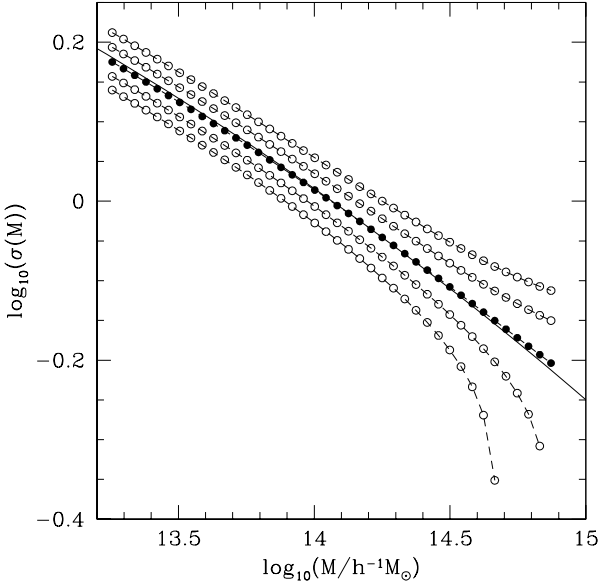}}
\caption{
Solutions for $\protect\sigma %
(M)$ obtained from the mass function of the $\Lambda $CDM GIF simulation by
the general fit of J01 fixing the values of $\Omega =0.3$, and varying $%
\protect\sigma _{8}$ in the values 0.8, 0.85, 0.9, 0.95 and 1.0. Filled
circles indicate the solution for $\protect\sigma _{8}=0.9$.
}
\label{fig8}
\end{figure}

\begin{figure}
{\epsfxsize=8.truecm \epsfysize=8.truecm 
\epsfbox[20 150 580 710]{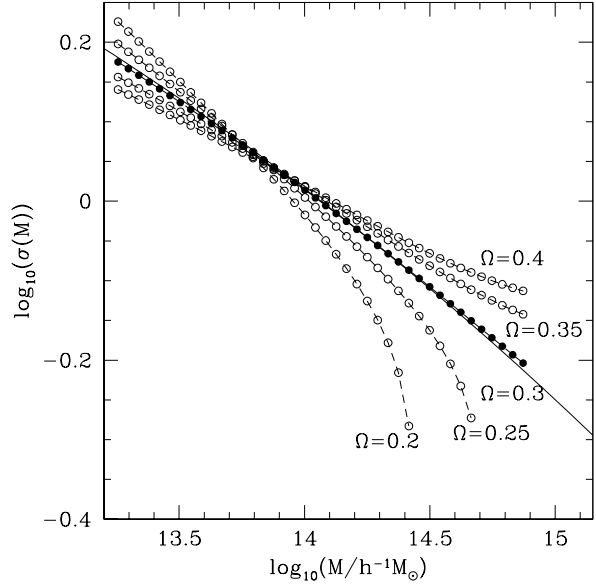}}
\caption{
Solutions for $\protect\sigma %
(M)$ obtained from the mass function of the $\Lambda $CDM GIF simulation using
the general fit of J01 fixing the values of $\protect\sigma _{8}=0.9$, and
varying $\Omega $ in the values 0.2, 0.25, 0.3, 0.35 and 0.4. Filled circles
indicate the solution for $\Omega =0.3$.
}
\label{fig9}
\end{figure}

\begin{figure}
{\epsfxsize=8.truecm \epsfysize=8.truecm 
\epsfbox[20 150 580 710]{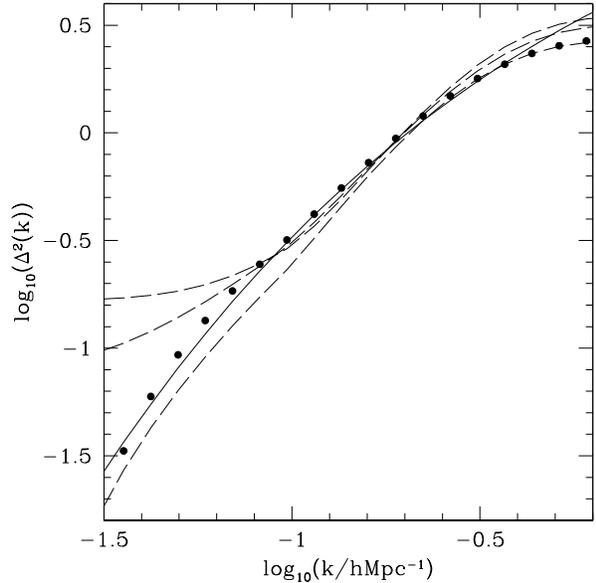}}
\caption{Solutions for $\Delta ^{2}(k)$ obtained aplying the LR technique 
to the solutions for $\sigma(R)$ from the $\Lambda$ CDM simulation by 
fixing $\Omega$ and varying $\sigma_8$ (see Figure 8). 
}
\label{fig10}
\end{figure}

\begin{figure}
{\epsfxsize=8.truecm \epsfysize=8.truecm 
\epsfbox[20 150 580 710]{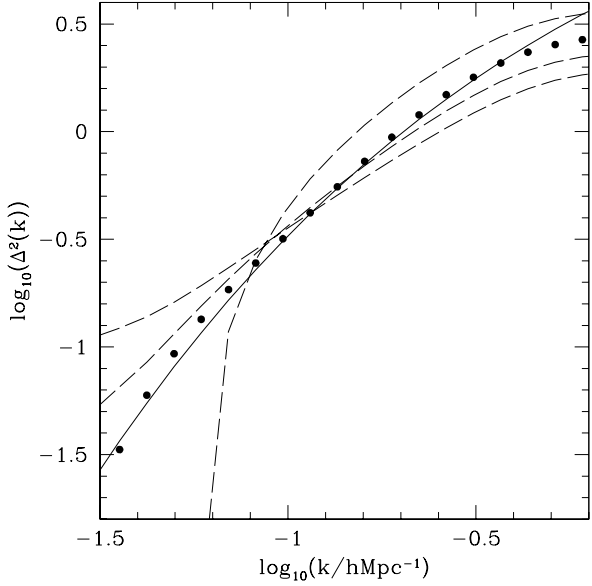}}
\caption{Solutions for $\Delta ^{2}(k)$ obtained aplying the LR technique 
to the solutions for $\sigma(R)$ from the $\Lambda$ CDM simulation by 
fixing $\sigma_8$ and varying $\Omega$ (see Figure 9). 
}
\label{fig11}
\end{figure}

The method for the determination of the power spectrum described in \S 3
requires the prior knowledge of the density parameter $\Omega $ and the 
rms mass fluctuations on some reference scale, 
for example $R=8h^{-1}$Mpc ($%
\sigma _{8})$. In this section we analyze the dependence of our
results on these parameters. In order to do so we calculate
the solution $\sigma (M)$ from the mass function of the $\Lambda $%
CDM simulation ($\Omega =0.3$, $\sigma _{8}=0.9$) fixing $\Omega =0.3$ and
setting $\sigma _{8}= 0.8, 0.85, 0.9, 0.95$, and $1.0$, and
fixing $\sigma _{8}=0.9,$ and varying $\Omega $ from 0.2, 0.25, 0.3 , 0.35
and 0.4. The results obtained in this way are plotted in Figures 8 and 9.

Figure 8 shows that except for the high mass end, the effect
of changing $\sigma _{8}$ in the solutions of $\sigma (M)$ is simply
a change in the overall normalization. On the other hand, the shape of the
solutions is affected in the high mass limit. As is shown below,
this effect propagates into the power spectrum causing differences in
the low $k$ end.

Figure 9 shows that the effect of changing the value of $\Omega $ can be
more serious, as the solutions obtained have different behaviors than that
observed in the $\Omega =0.3$ case. This will cause differences in
the obtained power spectrum in all scales.

In order to analyze the effect of these variations on the power 
spectrum we use the LR technique to obtain $\Delta ^{2}(k)$ from the 
$\sigma (M)$ functions plotted in Figures 8 and 9. In doing so we used 
the same value of $\lambda $ used in \S 4.3 when obtaining the power 
spectrum from this simulation. The results obtained this way are plotted 
in Figures 10 and 11.

Figure 10 shows that, as we pointed out previously, the effect 
arising from changing the value of $\sigma _{8}$ is important in the 
lower values of $k$, where it can produce considerable changes in
the solutions. For the higher values of $k$, the solutions
obtained are not very different for the one obtained for the correct value
of $\sigma _{8}$. This is the same effect observed in the analysis of 
$\sigma (M)$ where we only found a renormalization and not a change 
in shape.

The most important effect on the power spectrum, is that caused
by the use of a wrong value of $\Omega $. As it can be seen in Figure 11, the
solutions obtained differ notably between them, and show different
behavior in every scale. The reason for this is that 
the effect of changing $\Omega $
in $\sigma (M)$ is not a simple renormalization, but a change in the
slope of the solutions. This makes $\Omega $ the most important
parameter in the model.

\section{Comparison to the FKP method}

As a final test of our method we compare
the outcomes of the method presented in this work, and that of
FKP, as described in Zandivarez, Abadi \& Lambas (2001). 

 We applied both methods to the GIF simulations and the results 
are plotted in Figures 12 and 13 corresponding to the $\Lambda$CDM 
and $\tau $CDM models respectively, where we plot the power 
spectrum $P(k)$, output of the FKP technique, in logarithmic 
(upper panel) and linear scales (lower panel). In order to compare 
these results we have arbitrarily normalized the power spectrum
 derived from the FKP technique to that of the linear
mass power spectrum, which is the outcome of our method.

\begin{figure}
{\epsfxsize=8.truecm \epsfysize=8.truecm 
\epsfbox[20 150 580 710]{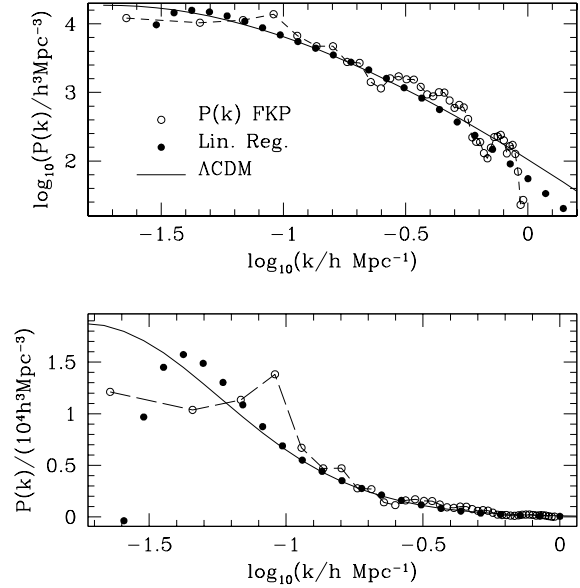}}
\caption{Results obtained for the power spectrum $P(k)$ of the $\Lambda$ 
CDM GIF simulation ussing the LR technique and the FKP method to the same 
group sample, with logarithmic (upper panel), and linear 
(lower panel) scales.
}
\label{fig12}
\end{figure}

\begin{figure}
{\epsfxsize=8.truecm \epsfysize=8.truecm 
\epsfbox[20 150 580 710]{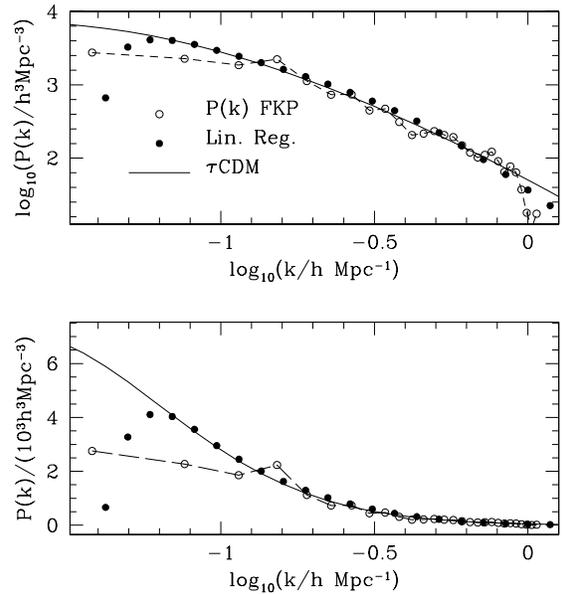}}
\caption{The same as Figure 12, but for the $\tau$ CDM GIF simulation.
}
\label{fig13}
\end{figure}

  By inspection to these figures, it is clear that FKP and our method 
yield similar results, showing the power and performance of our approach. 
We notice that
the solutions for $P(k)$ obtained from the mass function show a much 
better agreement with the power spectrum of the corresponding model than those 
obtained using the FKP technique. The range
of validity of our solutions must be kept in mind, since the apparent good 
description of $P(K)$ in the higher values of $k$ correspond to the region 
where LR gives a constant solution for $\Delta^2(k)$, which in turn gives 
$P(k)\alpha k^{-3}$, the approximate behavior of the power spectrum in
these scales. Even though LR gives a more 
accurate description of $P(k)$ in the range where it has been proven
to be applicable, 
the estimated power spectrum using the FKP technique 
works for higher values of $k$. As the range of validity of our 
solution depends on the mass range of the differential mass function,
the inclusion of low clusters masses will improve this situation. 

As it can be seen in the lower panels of Figures 12 and 13, 
the solution obtained by the use of LR 
gives a very accurate description of the power spectrum for the lower values of
 $k$, where the FKP method fails to give a satisfactory answer. Besides, the 
logarithmic bins in LR solution's sample these values of $k$ in a better way
 than the linear bins of the FKP technique, showing that our method is  
better in studying the power spectrum at these scales.

It is important to stress the fact that our method provides the linear
 power spectrum, while the solution of the FKP technique is affected by 
nonlinear effects. These are not important in the range of scales analyzed 
here, but this may be significant when objects of lower 
masses are included in the analysis.

A significant and important difference in our technique is that it naturally 
allows the use of information from different scales and objects, ranging from
poor to rich clusters of galaxies in an homogeneous way. On the other hand  
the FKP technique will give different results for subsets of systems of
galaxies with diferent masses, reflecting their relative bias with
respect to the underlying mass distribution. 
Furthermore, in our method it is desirable 
the inclusion of objects with a very broad range of masses, since this 
allows for a better determination of the power spectrum over a wider 
range of scales.

\section{Conclusions}

In this work we have developed a new method to determine the linear mass
power spectrum from the mass function of galaxy clusters based on the PS,
SMT and J01 formalisms. The rms fluctuation of mass $\sigma (M)$ is 
determined using the expression for the mass function in the form of a
differential equation. We then use different techniques from
integral equations theory to obtain the dimensionless power spectrum 
$\Delta ^{2}(k)$ from $\sigma (M)$.

The implementation of this method to the GIF $\tau $CDM and $\Lambda $CDM
simulations shows a high efficiency in recovering the correct shape and
amplitude of the power spectrum in a wide range of scales, showing a high 
performance and real practical utility. Although the different solutions 
for $\sigma (M)$, obtained from the simulations, show an overall agreement, 
it is clear that the J01 prescription gives a much better result than the PS 
or SMT formalism, showing that the J01 general fit gives an excellent 
description of the differential mass function of clusters identified in
the simulations.

In the theoretical case analyzed in \S 3.2 the best solution was
that obtained from the implementation of SVD, which recovers
the power spectrum in a wide range of scales. However, from the 
analysis of the results obtained from the simulations we conclude that the 
best technique to invert the integral equation in our problem is linear
regularization. This situation can change with the use of the full
covariance matrix and so, it must be analyzed without the assumption of a
diagonal $C_{S}$.

The comparison of the results obtained with our method with those obtained
by the standard FKP technique, shows a very good general agreement. 
Nevertheless, 
in the scales of validity of our solutions, our answer gives 
a much more accurate description
of the power spectrum than the ones obtained by the use of the FKP method. 
This is specially valid in 
the lower values of $k$ where the FKP solution fails to give the
correct answer. Further more, 
the logarithmic bins of our method sample these scales 
in a more efficient way than linear bins (as is the case in 
the FKP output). This produces important effects in the analysis of the 
power spectrum in large scales. The FKP technique works for higher 
values of $k$ than the LR solution.  
However, this can improve significantly if information
of $n(M)$ for lower masses is included in the analysis.

In order to apply our procedure to real data we need an estimate of the mass
function of galaxy clusters. Unfortunately, present estimates 
of this important statistical measure are not completely satisfactory  
since quoted values are significantly different and moreover they
are referred to different cluster mass definitions (see for instance
Bahcall \& Cen, 1993; Biviano et al., 1993; Girardi et al., \
1998; Girardi \& Giuricin, 2000; Reiprich \& B\"{o}hringer, 2001; 
Bahcall et al., 2002). 

As we have already pointed out, even in the case where errors in the 
individual masses are large, the overall effect in the mass 
function is not important when compared to other sources of 
uncertainties such as the use of different mass definitions.
Different methods for cluster mass estimates ( i.e. optical,
x-ray, gravitational lensing) are clearly correlated, but provide 
different values for the mass of the same cluster. Furthermore, 
it is not clear how these mass definitions are related to the ones 
used in numerical simulations ( i.e. FOF $b=0.2$, FOF $b=0.17$, SO 
$\kappa =200$, etc.) which also show differences among them ( see 
White, 2001). This is of extreme importance and is the principal source 
of uncertainties in our method. Given that observational cluster masses 
are estimated using one of the methods mentioned above, 
it may be incorrect to assume a fit to the mass function 
whose coefficients have been determined 
using FOF masses, such as the J01 formalism.

This indicates that the use of expressions derived from theoretical 
formalisms in the analysis of observational data ( a common practice 
in the literature) must not be made without a careful analysis of 
the relationship between the different mass definitions involved.

There are other statistics available in the literature, such as the
temperature function (Henry \& Arnaud, 1991, Markevitch, 1998 , 
Blanchard et al., 2000; Ikebe et al, 2001) of X-ray samples. The high 
correlation between temperature and mass suggested by numerical simulations 
(Evrard, 1990; Evrard, Metzler \& Navarro, 1996; Yoshikawa \& Suto, 2000, 
Mathiesen \& Evrard, 2001), has lead many authors to use temperature 
functions in confronting observations with theoretical predictions, instead 
of mass functions. 

The method described here can equally be applied to the 
temperature function of galaxy clusters, but, in doing so, we need a 
reliable mass-temperature relation. Unfortunately this transformation
has not been yet determined in an accurate way, and different $T-M$ 
relations give equally different results between them.

The observational determination of the mass function of 
galaxy clusters can improve in the next few years with the outcome 
of new mass estimates from gravitational lenses, or even from virial 
mass estimates obtained from the new large catalogs 
that will be soon available such as 2dF and SDSS. 

If the above discussion is properly addressed 
the methods described in this work can be very useful tools to  
provide new and reliable determinations of the linear mass power spectrum.

\section*{Acknowledgments}

We would like to thank Carlthon Baugh for reading the manuscript and his 
helpful comments and Ariel Zandivarez for providing us the 
calculations of the FKP power spectrum. We also want to thank the 
referee Fabio Governato for his constructive suggestions. AGS also 
thanks Mario Abadi for his kind help. This work was 
partially supported by the Concejo Nacional de Investigaciones 
Cient\'{\i}ficas y Tecnol\'ogicas (CONICET), the Secretar\'{\i}a de 
Ciencia y T\'ecnica (UNC) and the Agencia C\'ordoba Ciencia.

\appendix

\section{Obtaining $\protect\sigma (M)$ from the 
PS, SMT and J01 formalisms}

The expression for the mass function in the PS, SMT and J01 formalisms can
be re-interpreted as differential equations to obtain $\sigma (M)$ from $%
n(M) $. If we use the PS formalism then the mass function is given by:

\[
n(M,z)dM=\sqrt{\frac{2}{\pi }}\frac{\bar{\rho}}{M^{2}}\frac{\delta _{c}}{%
\sigma (M,z)}\left| \frac{d\ln (\sigma )}{d\ln (M)}\right| e^{-\left( \frac{%
\delta _{c}^{2}}{2\sigma ^{2}(M,z)}\right) }dM 
\]
that can be written as 
\[
Mn(M)dM=-\sqrt{\frac{2}{\pi }}\frac{\bar{\rho}}{\sigma ^{2}}e^{-\left( \frac{%
\delta _{c}^{2}}{2\sigma ^{2}}\right) }\frac{d\sigma }{dM}dM 
\]
where all dependence on $\sigma $ is in the right hand side. Integrating
in both sides with respect to $M$ between a mass $M$ and $M_{8}=$ $%
\frac{4\pi }{3}\bar{\rho}(8h^{-1}Mpc)^{3}$ corresponding to a sphere of
uniform density and radius $R=8h^{-1}Mpc$, we get

\begin{equation}
G(M)\equiv \int_{M}^{M_{8}}\grave{M}n(\grave{M})d\grave{M}=\bar{\rho}\frac{2%
}{\sqrt{\pi }}\left[ \int_{\mu _{M}}^{\mu _{8}}e^{-\mu ^{2}}d\mu \right]
\label{uno}
\end{equation}
where $\mu =\frac{\delta _{c}}{\sqrt{2}\sigma (M)}$ and $\mu _{8}=\frac{%
\delta _{c}}{\sqrt{2}\sigma _{8}}$.

Using the definition of the error function

\[
{\rm erf}(x)=\frac{2}{\sqrt{\pi }}\int_{0}^{x}e^{-\mu ^{2}}d\mu 
\]
and using the value of $\sigma _{8}$ as a constant of integration, (\ref{uno}%
) can be expressed as

\[
G(M)=\bar{\rho}\left[ {\rm erf}(\frac{\delta _{c}}{\sqrt{2}\sigma _{8}})-%
{\rm erf}(\frac{\delta _{c}}{\sqrt{2}\sigma (M)})\right] 
\]
and then we obtain the expression for $\sigma (M)$ given in equation (\ref
{sigmadeps})

\[
\sigma (M)=\frac{\delta _{c}}{\sqrt{2}{\rm erf}^{-1}\left[ {\rm erf}(\frac{%
\delta _{c}}{\sqrt{2}\sigma _{8}})-\frac{G(M)}{\bar{\rho}}\right] } 
\]

If we assume the validity of the SMT formalism then the differential equation
for $\sigma (M)$ is

\[
Mn(M)dM=-\sqrt{\frac{2}{\pi }}A\bar{\rho}\frac{\sqrt{a}\delta _{c}}{\sigma
^{2}}\left[ 1+\left( \frac{\sigma }{\sqrt{a}\delta _{c}}\right) ^{2p}\right]
e^{-\left( \frac{a\delta _{c}^{2}}{2\sigma ^{2}}\right) }\frac{d\sigma }{dM}%
dM 
\]

This is a little more complicated than the corresponding to the PS formalism
but can be solved in an analogous way. Integrating between $M$ and $M_{8}$
we get 

\begin{equation}
G(M)=A\bar{\rho}\frac{2}{\sqrt{\pi }}\left[ \int_{\mu _{M}}^{\mu
_{8}}e^{-\mu ^{2}}d\mu +2^{-p}\int_{\mu _{M}}^{\mu _{8}}\mu ^{-2p}e^{-\mu
^{2}}d\mu \right]  \label{resulprevsmt}
\end{equation}
where $G(M)$ is given by (\ref{uno}) and now $\mu =\sqrt{\frac{a}{2}}\frac{%
\delta _{c}}{\sigma (M)}.$

The first term of the right hand side in (\ref{resulprevsmt}) is the same
than in the PS formalism and is equal to ${\rm erf}(\mu )\mid _{\mu
_{M}}^{\mu _{8}}$, the difference in this case is given by the second term%
\textit{. }If we use the definition of the \textit{incomplete gamma
function}

\[
P(a,x)=\frac{1}{\Gamma (a)}\int_{0}^{x}t^{a-1}e^{-t}dt\qquad a>0 
\]
we can see easily that

\begin{eqnarray*}
\int_{0}^{\mu _{M}}\mu ^{-2p}e^{-\mu ^{2}}d\mu &=&\frac{1}{2}\int_{0}^{\mu
_{M}^{2}}t^{-\left( p+\frac{1}{2}\right) }e^{-t}dt \\
&=&\frac{1}{2}\Gamma \left( \frac{1}{2}-p\right) P\left( \frac{1}{2}-p,\mu
_{M}^{2}\right)
\end{eqnarray*}
Then, if we define the function

\[
\Phi (x)={\rm erf}(x)+\frac{2^{-p}}{\sqrt{\pi }}\Gamma \left( \frac{1}{2}%
-p\right) P\left( \frac{1}{2}-p,x^{2}\right) 
\]
we can write (\ref{resulprevsmt}) as

\[
G(M)=A\bar{\rho}\left[ \Phi (\frac{\delta _{c}}{\sqrt{2}\sigma _{8}})-\Phi (%
\frac{\delta _{c}}{\sqrt{2}\sigma (M)})\right] 
\]
from where we obtain the expression for $\sigma (M)$ in the SMT formalism given
in equation (\ref{sigmadesmt})

\[
\sigma (M)=\frac{\delta _{c}}{\sqrt{2}\Phi ^{-1}\left[ \Phi (\frac{\delta
_{c}}{\sqrt{2}\sigma _{8}})-\frac{G(M)}{\bar{\rho}A}\right] } 
\]

It is possible to consider that 
the J01 mass function gives a more accurate description of
the real behavior of the mass function.  If this is the case,
then we use the expression

\[
n(M,z)dM=\frac{A\bar{\rho}(z)}{M^{2}}\frac{d\ln (\sigma ^{-1})}{d\ln (M)}%
e^{-\left( \left| \ln (\sigma ^{-1}+B\right| ^{\varepsilon }\right) }dM 
\]
which must be re-interpreted as a differential equation for $\sigma (M)$ as a
function of $n(M,z)$. This expression can be written as 
\[
Mn(M)dM=A\bar{\rho}e^{-\left( \left| S+B\right| ^{\varepsilon }\right) }%
\frac{dS}{dM}dM 
\]
where $S\equiv \ln (\sigma ^{-1})=-\ln (\sigma ).$ Integrating in both sides
with respect to $M$ between $M$ and $M_{8}$ and keeping the definition of $%
G(M)$, we get 
\[
G(M)=A\bar{\rho}\left( \Psi (S_{8})-\Psi (S)\right) 
\]
where 
\[
\Psi (S)=\int_{0}^{S}e^{-\left| x+B\right| ^{\varepsilon }}dx 
\]
Then we can write
\[
S=\Psi ^{-1}\left[ \Psi (S_{8})-\frac{G(M)}{A\bar{\rho}}\right] 
\]
from where we get the following expression for $\sigma (M)$%
\[
\sigma (M)=\exp \left\{ -\Psi ^{-1}\left[ \Psi (\ln \sigma _{8}^{-1})-\frac{%
G(M)}{A\bar{\rho}}\right] \right\} 
\]
which is given in equation (\ref{sigmadej01}).

\end{document}